\documentclass{article}%
\usepackage{graphics}
\usepackage{graphicx}
\usepackage{amsmath}
\usepackage{amsfonts}
\usepackage{amssymb}%
\newtheorem{theorem}{Theorem}

\newtheorem{definition}[theorem]{Definition}
\newtheorem{example}[theorem]{Example}

\newtheorem{proposition}[theorem]{Proposition}
\newtheorem{remark}[theorem]{Remark}

\newenvironment{proof}[1][Proof]{\textbf{#1.} }{\ \rule{0.5em}{0.5em}}

\begin{document}

\date{}
\title{Meta-dynamical adaptive systems and their applications to a fractal algorithm
and a biological model}
\author{Emmanuel MOULAY\thanks{E-mail:
Emmanuel.Moulay@irccyn.ec-nantes.fr}\\IRCCyN (UMR-CNRS 6597)\\ 1
rue de la No\"{e}\\ B.P. 92 101 \\44321 Nantes Cedex 03, France
\\\\ Marc
BAGUELIN\thanks{E-mail: mb556@cam.ac.uk}\\Department of Zoology\\ University of Cambridge\\
Downing Street\\ Cambridge, CB2 3EJ, UK}

\maketitle

\begin{abstract}
In this article, one defines two models of adaptive systems: the
\emph{meta-dynamical adaptive system }using the notion of Kalman
dynamical systems and the \emph{adaptive differential equations}
using the notion of \emph{variable dimension spaces}. This concept
of variable dimension spaces relates the notion of spaces to the
notion of dimensions. First, a computational model of the Douady's
Rabbit fractal is obtained by using the meta-dynamical adaptive
system concept. Then, we focus on a defense-attack biological
model described by our two formalisms.\newline \textbf{Keywords:
}dynamical systems, adaptive systems, biological systems, fractal
algorithm.\newline \textbf{MSC: } 93A05; 37F50; 92D15; 92D25
\end{abstract}

\section{Introduction}

In the two last decades, there has been much interest in the study and
formalization of complex adaptive systems (see \cite{Waldrop92} and
\cite{Gell-Mann94}). Many different approaches have been proposed: artificial
chemistries, evolutionary formalism, cellular automata. However, from a more
theoretical viewpoint, few mathematical formalisms exist for adaptive systems.
Though we may cite the chapter \textquotedblleft Categorical System
Theory\textquotedblright\ proposed by A.H. Louie in \cite{Rosen85} who discuss
the relationship between natural and formal systems, most of the studies are
mainly simulation-based (see \cite{Gell-Mann94} and \cite{Holland92}). With
the recent evolution of physics and biology, a general mathematical formalism
for adaptive systems would be very useful. Actually, behind their apparent
heterogeneity, most of the adaptive systems share one important feature: they
are dynamical objects whose structures are sometimes modified by a top level
automation-like rule. On the basis of this observation, we have built a
two-level formalism that helps us to design an algorithm for fractals and a
biological model of co-evolution in a bacterium-phage system.

The main goal of this article is to present a concept of adaptive systems,
general enough to be used in different fields (mathematics, physics or
biology) and to apply this framework to a fractal implementation and a
biological system. Besides, a new concept of spaces is developed. Since the
second part of the 19$^{th}$ century, various definitions of dimensions, in
particular those developed by Cantor and Peano, have appeared (see
\cite{Mandelbrot82}). These new definitions lead to the concept of a fractal
model and with it, a global view of spaces (see \cite{Mandelbrot80}). In order
to study adaptive systems, we will use a sort of \textquotedblleft
adaptive\textquotedblright\ space called $\emph{variable}$ $\emph{dimension}$
$\emph{space}$ whose dimension changes. This new kind of space brings the
notion of space and the notion of dimension together.\newline The paper is
organized as follows. In section \ref{SecMAS}, a very general adaptive system
concept called \emph{meta-dynamical adaptive system} is given by using an
extension of Kalman systems. Section \ref{SecDRF} is dedicated to the Douady's
Rabbit fractal implemented as a meta-dynamical adaptive system. In section
\ref{SecADE}, the concept of variable dimension space leads to a special model
of adaptive system: the \emph{adaptive differential equation}. Finally, we use
our concepts of adaptive systems to describe a biological example of a
defense-attack model in section \ref{Sec BIO}.

\section{Meta-dynamical adaptive system\label{SecMAS}}

It is possible to give a general definition of adaptive systems by using the
concept of \emph{meta-dynamical adaptive system }developed by one of the
author during his Ph.D. (\cite{Baguelin03,BaguelinECC03}). But, in order to be
able to present our formal approach, we will have a look at the main ideas
behind the formalization of the concept of dynamical system by Kalman in
\cite{Kalman69}. The aim of Kalman's approach is to show that some very common
mathematical structures plus a few axioms can provide a very general framework
where the notion of dynamical systems (of all kinds) is captured. Time is
modelled as an ordered subset of the reals (to cover both the continuous and
discrete paradigms). The important objects are the state set (the variables
characterizing the system) and the transition function. The transition
function defines the trajectory in the state set starting from an initial
state. Only a few axioms are required to characterize these objects and allow
them to form a \textquotedblleft dynamical system\textquotedblright. Among the
more important ones are direction of time, consistency, composition property
and causality. Let us recall the fundamental definition of Kalman (see
\cite{Kalman69}). \newline A \emph{dynamical system respecting Kalman axioms}
is defined by the following axioms:

\begin{enumerate}
\item There is a given time set $T$, a state set $X$, a set of input values
$U$, a set of acceptable input functions $\Omega=\{\omega:T\rightarrow U\}$, a
set of output values $Y$, and a set of output functions $\Gamma=\left\{
\gamma:T\rightarrow Y\right\}  $,

\item (Direction of time) $T$ is an ordered subset of the reals\footnote{Such
a general definition could include exotic sets such as fractal cantor sets,
for instance, in practice, the sets used are part of $\mathbb{R}$ or
$\mathbb{N}$.},

\item The input space $\Omega$ satisfies the following conditions:

\begin{enumerate}
\item (Nontriviality) $\Omega$ is nonempty,

\item (Concatenation of inputs) An input segment $\omega_{\left]  t_{1}%
,t_{2}\right]  }$ is $\omega\in\Omega$ restricted to $\left]  t_{1}%
,t_{2}\right]  \cap T$. If $\omega,\omega^{\prime}\in\Omega$ and $t_{1}%
<t_{2}<t_{3}$, there is an $\omega^{\prime\prime}\in\Omega$ such that
$\omega_{\left]  t_{1},t_{2}\right]  }^{\prime\prime}=\omega_{\left]
t_{1},t_{2}\right]  }$ and $\omega_{\left]  t_{2},t_{3}\right]  }%
^{\prime\prime}=\omega_{\left]  t_{2},t_{3}\right]  }^{\prime}$,
\end{enumerate}

\item There is given a state-transition function%
\[
\varphi:T\times T\times X\times\Omega\rightarrow X.
\]
whose value is the state $x\left(  t\right)  =\varphi\left(  t,\tau,x_{\tau
},\omega\right)  \in X$ resulting at time $t\in T$ from the initial state
$x_{\tau}=x\left(  \tau\right)  \in X$ at initial time $\tau\in T$ under the
action of the input $\omega\in\Omega$. $\varphi$ has the following properties:

\begin{enumerate}
\item (Direction of time) $\varphi$ is defined for all $t\geq\tau$ but not
necessarily for all $t<\tau$.

\item (Consistency) $\varphi\left(  t,t,x,\omega\right)  =x$ for all $t\in T$,
all $x\in X$ and all $\omega\in\Omega$.

\item (Composition property) For any $t_{1}<t_{2}<t_{3}$, we have%
\[
\varphi\left(  t_{3},t_{1},x,\omega\right)  =\varphi\left(  t_{3}%
,t_{2},\varphi\left(  t_{2},t_{1},x,\omega\right)  ,\omega\right)
\]
for all $x\in X$ and all $\omega\in\Omega$.

\item (Causality) If $\omega,\omega^{\prime}\in\Omega$ and $\omega_{\left]
\tau,t\right]  }=\omega_{\left]  \tau,t\right]  }^{\prime}$ then%
\[
\varphi\left(  t,\tau,x,\omega\right)  =\varphi\left(  t,\tau,x,\omega
^{\prime}\right)  .
\]

\end{enumerate}

\item There is given a readout map $\eta:T\times X\rightarrow Y$ which defines
the output $y\left(  t\right)  =\eta\left(  t,x\left(  t\right)  \right)  $.
The map $\left]  \tau,t\right]  \rightarrow Y$ given by $\sigma\mapsto
\eta\left(  \sigma,\varphi\left(  \sigma,\tau,x,\omega\right)  \right)  $,
$\sigma\in\left]  \tau,t\right]  $ is an output segment, that is, the
restriction $\gamma_{\left]  \tau,t\right]  }$ of some $\gamma\in\Gamma$ to
$\left]  \tau,t\right]  $.
\end{enumerate}

A dynamical system is referred as $\Sigma=\left\{  T,X,U,\Omega,Y,\Gamma
,\varphi,\eta\right\}  $.

Now, inspired by the definition of dynamical systems of Kalman, we propose a
formalization of our meta-dynamical adaptive system.

\begin{definition}
\label{MAS}A \emph{meta-dynamical adaptive system }$\mathcal{M}$ is a
composite mathematical concept defined by the following axioms:

\begin{enumerate}
\item \textbf{The dynamical level}: the suitably indexed set
\[
\Sigma_{i,j}=\{T_{0},X_{i},U,\Omega,Y,\Gamma,\varphi_{i,j}\}
\]
is a dynamical system respecting Kalman axioms for all $\left(  i,j\right)
\in I\times J$ where $\left\{  \varphi_{i,j}\right\}  _{j\in J}$ are
transition functions on state set $X_{i}$:%
\[
\varphi_{i,j}:T_{0}\times T_{0}\times X_{i}\times\Omega\rightarrow X_{i}.
\]

\item \textbf{The meta-dynamical level}: let $X=\bigcup\limits_{i\in I}X_{i}$
be the set of all the possible states of the system and $D=\{\varphi
_{i,j}\}_{\left(  i,j\right)  \in I\times J}$ be the set of all possible
transition functions, then there exists a meta-dynamical time $T_{1}$ and a
meta-dynamical rule\footnote{It is possible to consider input values at the
metadynamical level, then $\Phi$ is defined as follows
\[
\Phi:T_{1}\times X\times D\times\Omega_{1}\rightarrow X\times D
\]
where $\Omega_{1}=\left\{  \omega:T_{1}\rightarrow U_{1}\right\}  $ is the set
of acceptable \textquotedblleft meta\textquotedblright-input functions.}
\[
\Phi:T_{1}\times X\times D\rightarrow X\times D
\]
such that:

\begin{enumerate}
\item (Temporal hierarchy) $T_{1}\subseteq T_{0}$,

\item (Coherence states/transitions) If $\Phi(t,x_{1},\varphi_{i_{1},j_{1}%
})=(x_{2},\varphi_{i_{2},j_{2}})$ for $t\in T_{1}$, then $x_{1}\in X_{i_{1}%
}\Rightarrow x_{2}\in X_{i_{2}}$.
\end{enumerate}

If $t\in T_{1}$ and $\Phi(t,x,\varphi)=(x,\varphi)$, $\Phi$ is said to be
\emph{mute} at $(t,x,\varphi)$ else $(t,x,\varphi)$ is a \emph{commutation
point}.

\item \textbf{Evolution rule between dynamical and meta-dynamical levels}: let
$x_{t}=\varphi(t,t_{1},x_{1},\omega)$ with $\omega$ an input function, there
exists a \textquotedblleft meta\textquotedblright-transition function
\[
\Psi:T_{1}\times T_{1}\times X\times D\times\Omega\rightarrow X\times D
\]
such that:

\begin{enumerate}
\item (Dynamical phase) If $\Phi$ is mute on $(t,x_{t},\varphi)$ for all
$t\in\lbrack t_{1},t_{2}[\cap T_{1}$, then $\Psi$ is defined between $t_{1}$
and $t_{2}$ and

\begin{enumerate}
\item if $t_{2}\notin T_{1}$, then
\[
\Psi(t_{2},t_{1},x_{1},\varphi,\omega)=(x_{t_{2}},\varphi),
\]

\item else $t_{2}\in T_{1}$ and
\[
\Psi(t_{2},t_{1},x_{1},\varphi,\omega)=\Phi(x_{t_{2}},\varphi).
\]

\end{enumerate}

\item (Concatenation rule) If there exists $t_{2}\in]t_{1},t_{3}[$ such that
$\Psi$ is defined between $t_{1}$ and $t_{2}$ and between $t_{2}$ and $t_{3}$
then $\Psi$ is defined between $t_{1}$ and $t_{3}$ and
\[
\Psi(t_{3},t_{1},x_{1},\varphi,\omega)=\Psi(t_{3},t_{2},\Psi(t_{2},t_{1}%
,x_{1},\varphi,\omega),\omega).
\]

\item (Stopping rule) $\Psi$ is defined between $t_{1}$ and $t_{2}$ if,
respecting previous axioms, there is only a finite number of commutation
points in $[t_{1},t_{2}]$.
\end{enumerate}
\end{enumerate}
\end{definition}

The meta-dynamical rule in point $2)$ can operate at instants for which the
system is defined and not necessarily at all of them (see point $2.(a)$);
actually a higher level is usually slower. Moreover, as the meta-dynamical
rule can change the state and the transition function, we have to consider
that both match well: the resulting state has to belong to the state set on
which the new transition function operates (point $2.(b)$). In point $3.$ we
describe how dynamics and the meta-dynamical rule combine together to make the
system change with time.

To take up notions mainly used in social science, our dynamical rule can be
seen as a heterarchical level and our meta-dynamical level as a hierarchical
level. Let us recall that a \emph{heterarchy} is a network of elements which
share the same \textquotedblleft horizontal\textquotedblright\ position level
in a \emph{hierarchy}. Each level in a hierarchical system is composed of a
heterarchy which contains its constituent elements (see \cite{Jen03}).

\begin{remark}
In the second case of $\left(  3.a\right)  $, when $t_{2}\in T_{1}$, we
deliberately consider that $\left(  t_{2},\Phi\left(  x_{t_{2}},\varphi
\right)  \right)  $ is mute. We do not consider the case where the
meta-dynamics would \textquotedblleft rebound\textquotedblright\ and have
several commutations at the same time. If the system has several commutations,
it is always possible to consider this set of commutation as one, with the
final state of the last commutation (provided that, we know from $3.(c)$ that
the number of commutations is finite). Axiom $3.(c)$ is to avoid Zeno-style
system with an infinite number of commutations in a finite amount of time
(e.g. with a quantity $x(t)=\sin\left(  \frac{2\pi}{1-t}\right)  $ with a
commutation each time $x(t)=0$ on $\left[  1-\epsilon,1+\epsilon\right]
$).\linebreak
\end{remark}

\begin{remark}
The index set $J$ is associated with the change of dynamics in the same state
space. We recognize here the framework of hybrid systems (see \cite{Zeigler00}).
\end{remark}

\begin{remark}
It is also important to differentiate between the continuous or discrete
dynamics and the continuous or discrete meta-dynamics because the confusion is
easy. The first case is well known. The second one shows the difference
between a meta-dynamical time $T_{1}$ which is continuous (see section
\ref{SecDRF}) and a meta-dynamical time which is discrete (see section
\ref{Sec BIO}).
\end{remark}

\begin{figure}[h]
\centering
\includegraphics[width=12cm]{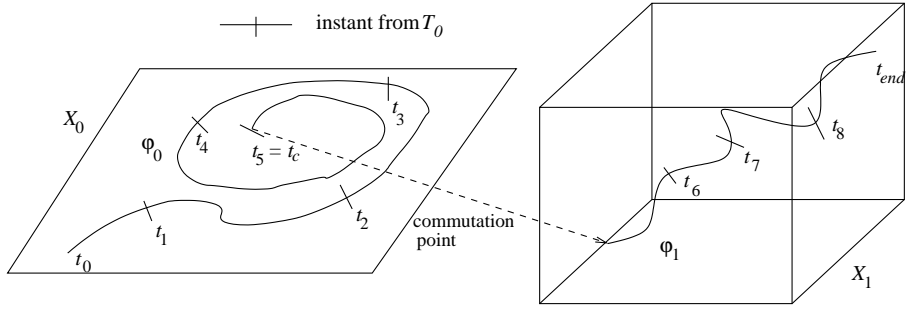}\caption{Illustration of some
axioms of a metadynamical system}%
\label{commu}%
\end{figure}

\begin{example}
\label{Ex MAS}On Fig. \ref{commu}, we can see a meta-dynamical system in
action. At $t_{5}=t_{c}$, we pass from a $2-$dimensional state set to a
$3-$dimensional one. The dynamics are continuous, so $T_{0}=[t_{0},t_{end}]$
and the meta-dynamical time set $T_{1}$ is discrete: $T_{1}=\{t_{1}%
,t_{2},t_{3},\ldots,t_{8}\}$. Since the meta-dynamics is mute on $t_{1}$,
$t_{2}$, $t_{3}$ and $t_{4}$, the evolution from $t_{0}$ to $t_{5}$ is a
purely dynamical phase, ended by a commutation (Axiom $3.(a)$). Evolution from
$t_{5}$ to $t_{end}$ is also purely dynamical (the meta-dynamics is mute on
$t_{6}$, $t_{7}$ and $t_{8}$. The junction between the two of them is made by
using the concatenation rule. As there is only one commutation, the system is defined.
\end{example}

Now, it is possible to expand some classical notions of dynamical system, as
the notion of trajectory.

\begin{definition}
Let $\mathcal{M}$ be a meta-dynamical adaptive system with the meta-transition
function $\Psi$, if for $t\in T_{1}$,
\[
\Psi(t,t_{0},x_{0},\varphi_{0},\omega)=\Phi(x_{t},\varphi_{t})
\]
with $x_{t}\in X_{t}$ then $\left(  X_{t},\varphi_{t}\right)  $ is a
$\emph{meta-state}$ of $\mathcal{M}$ in $t$.

The sequence $\left(  X_{t_{i}},\varphi_{t_{i}}\right)  _{0\leq i\leq n,0\leq
n\leq+\infty}$ of meta-state of $\mathcal{M}$ such that for all $t\in\left[
t_{i},t_{i+1}\right[  $, $\left(  X_{t},\varphi_{t}\right)  =\left(  X_{t_{i}%
},\varphi_{t_{i}}\right)  $ and $\left(  X_{t_{i}},\varphi_{t_{i}}\right)
\neq\left(  X_{t_{i+1}},\varphi_{t_{i+1}}\right)  $ is called $\emph{meta-}%
$\emph{orbit} of $\mathcal{M}$.

If $n<+\infty$, $\left(  X_{t_{n}},\varphi_{t_{n}}\right)  $ is said to be an
\emph{absorbent} meta-state.

The sequence $\left(  t_{i},X_{t_{i}},\varphi_{t_{i}}\right)  _{0\leq i\leq
n,0\leq n\leq+\infty}$ satisfying the same conditions is called a
\emph{trajectory} of $\mathcal{M}$.
\end{definition}

One can extend definition \ref{MAS} to stochastic systems.

\begin{definition}
\label{SMAS}Let us consider the dynamical rule of definition \ref{MAS} with
the following meta-dynamical rule
\[
p\Phi:T_{1}\times X\times D\rightarrow\left(  X\times D\rightarrow\left[
0,1\right]  \right)
\]
where $p\Phi$ is a probability distribution on $X\times D$ which represents
the probability
\[
p\Phi\left(  t,x,\varphi\right)  \cdot\left(  x_{f},\varphi_{f}\right)
\]
that $\left(  x,\varphi\right)  $ becomes $\left(  x_{f},\varphi_{f}\right)  $
at $t$. The stochastic evolution rule between dynamical and meta-dynamical
levels is given by
\[
p\Psi:T_{1}\times T_{1}\times X\times D\times\Omega\rightarrow\left(  X\times
D\rightarrow\left[  0,1\right]  \right)
\]
where $p\Psi$ is a probability distribution on $X\times D$ which represents
the probability
\[
p\Psi\left(  t_{1},t_{2},x,\varphi,\omega\right)  \cdot\left(  x_{f}%
,\varphi_{f}\right)
\]
that $\left(  x,\varphi\right)  $ becomes $\left(  x_{f},\varphi_{f}\right)  $
between $t_{1}$ and $t_{2}$ with the input function $\omega$. The properties
of $p\Phi$ and $p\Psi$ are the same as $\Phi$ and $\Psi$ given in \ref{MAS}.
Such a system is called a \emph{stochastic meta-dynamical adaptive system}.
\end{definition}

The interest of definition \ref{SMAS} is its generality. It is for example
well adapted to model a large number of complex adaptive systems, in
particular\ the biological model we develop in section \ref{Sec BIO}.

\section{Algorithm for the Douady's Rabbit fractal\label{SecDRF}}

The goal of this example is to give an algorithm based on definition \ref{MAS}
allowing to describe the Douady's Rabbit Fractal. In this example, the set $I$
and thus the family $\left\{  X_{i}\right\}  _{i\in I}$ of definition
\ref{MAS} is not countable. The Riemann Sphere $\mathbb{S}^{2}$ is mapped
one-to-one onto the extended complex plane $\mathbb{C}_{\infty}=\mathbb{C\cup
}\left\{  \infty\right\}  $ by stereographic projection. Let us recall some
definitions. Let $\mathbb{K}=\mathbb{R}$ or $\mathbb{C}$ and $\mathcal{B}$ be
the unit ball of $\mathbb{K}$. If $f:\mathbb{K}\rightarrow\mathbb{K}$ is a
function, $x\in X$ is $\emph{periodic}$ of period $n\in\mathbb{N}$ if
$f^{n}(x)=x$ and for all $k\in\{1,..,n-1\}$, $f^{k}(x)\neq x$. For $x$
periodic of period $n$, the \emph{cycle} $O(x)$ is the set
\[
\mathcal{O}(x)=\left\{  x,f(x),...,f^{n-1}(x)\right\}
\]
and its cardinal is $n$. Moreover, if $f$ is differentiable at $x$, $x$ is
\emph{stable}, $\emph{quasi}$ $\emph{stable}$ or \emph{unstable} if
$\left\vert \left(  f^{n}\right)  ^{\prime}(x)\right\vert <1$, $\left\vert
\left(  f^{n}\right)  ^{\prime}(x)\right\vert =1$ or $\left\vert \left(
f^{n}\right)  ^{\prime}(x)\right\vert >1$. If $x$ is stable, $x$ is
\emph{attractive} if there exists an interval $V$ strictly containing $x$ so
that for all $x\in V$%
\[
f^{n}(x^{\prime})\underset{n\rightarrow+\infty}{\rightarrow}x.
\]
Now, let us recall the definition of the Julia set (see \cite{Alexander94}).
Let $R$ be a non-constant rational function on $\mathbb{S}^{2}$. The
\emph{Fatou set} of $R$ is the maximal open subset of $\mathbb{S}^{2}$ on
which $\left\{  R^{n}\right\}  _{n\in\mathbb{N}}$ is equicontinuous where
$R^{n}=R\circ\ldots\circ R$. The \emph{Julia set }$J_{R}$ of $R$ is the
complement of the Fatou set on $\mathbb{S}^{2}$. The \emph{filled in Julia
set} $K_{R}$ of a function $R$ is all the points which are not attracted to
the super-attracting fixed point at infinity, that is
\[
K_{R}=\left\{  z\in\mathbb{C}:R^{n}\left(  z\right)  \rightarrow
\infty\right\}  .
\]
This closed set includes the Julia set as its boundary, $J_{R}=\partial K_{R}%
$. The \emph{escape set} $I_{R}$ of a function $R$ is all the points that
\textquotedblleft escape\textquotedblright\ to infinity, that is
\[
I_{R}=\left\{  z\in\mathbb{C}_{\infty}:R^{n}\left(  z\right)  \rightarrow
\infty\right\}  .
\]
If $R$ is a polynomial of degree $2$, $J_{R}$ is called a quadratic Julia set.
The following result can be found in \cite{Brolin65}: if $P$ is a polynomial
of degree $d\geq2$, then $J_{P}$ is closed and dense within itself.\newline
Here, we are interested in quadratic Julia sets with
\begin{equation}
P_{c}(z)=z^{2}+c \label{Poly}%
\end{equation}
where $c\in\mathbb{C}$. For small values of $c$, the Julia set is distorted by
varying degrees from the unit circle, in these cases the Julia set has an
infinite length. For large values of $c$, the Julia set becomes an infinite
set of totally disconnected points, often said to be dust like (in the sense
of Cantor). In the quadratic case there are only these two possible -connected
and disconnected- types. The \emph{Mandelbrot set} is the space containing the
value $c$ for which the associated Julia set is connected. It is generated by
the quadratic sequence
\begin{equation}
\left\{
\begin{array}
[c]{l}%
z_{n+1}=z_{n}^{2}+c\\
z_{0}=c
\end{array}
\right.  \label{Julia}%
\end{equation}
A complex point $z=a+ib\in\mathbb{C}$ will be denoted by $(a;b)$. The Douady's
Rabbit fractal is a Mandelbrot set with $c=(-\frac{3}{2}+\frac{1}%
{2}(a+b);\frac{\sqrt{3}}{2}(a-b))$ where $a=\sqrt[3]{\frac{25+\sqrt{621}}{54}%
}$ and $b=\sqrt[3]{\frac{25-\sqrt{621}}{54}}$, so $c\simeq\left(
-0.12256;0.744862\right)  $ (see \cite{Douady84}). As the points get closer to
the Douady's Rabbit fractal, the speed of convergence becomes slower. With the
choice of $c$, $\left(  c^{2}+c\right)  ^{2}+c=0$ so the origin is an
attractive cycle\footnote{Let $\mathbb{K}=\mathbb{R}$ or $\mathbb{C}$,
$f:\mathbb{K}\rightarrow\mathbb{K}$ be function, $x\in X$ is $\emph{periodic}$
of period $n\in\mathbb{N}$ if $f^{n}(x)=x$ and for all $k\in\{1,..,n-1\}$,
$f^{k}(x)\neq x$. For $x$ periodic of period $n$, the \emph{cycle} $O(x)$ is
the set
\[
\mathcal{O}(x)=\left\{  x,f(x),...,f^{n-1}(x)\right\}
\]
and its cardinal is $n$. Moreover, if $f$ is differentiable at $x$, $x$ is
\emph{stable}, $\emph{quasi}$ $\emph{stable}$ or \emph{unstable} if
$\left\vert \left(  f^{n}\right)  ^{\prime}(x)\right\vert <1$, $\left\vert
\left(  f^{n}\right)  ^{\prime}(x)\right\vert =1$ or $\left\vert \left(
f^{n}\right)  ^{\prime}(x)\right\vert >1$. If $x$ is stable, $x$ is
\emph{attractive} if there exists an interval $V$ strictly containing $x$ so
that for all $x\in V$%
\[
f^{n}(x^{\prime})\underset{n\rightarrow+\infty}{\rightarrow}x.
\]
\par
{}} of\ period $3$ of $P_{c}$
\[
\mathcal{O}(0)=\left\{  0,c,c^{2}+c\right\}  .
\]
The boundary points move chaotically. Thus, the idea is to change\ the points
by using the speed of convergence as an adaptive value. The principle of our
algorithm is to have a set of points evolved to the fractal boundary. For
this, one gives a weight to each point. This weight varies with a dynamics
which \textquotedblleft rewards\textquotedblright\ the most adapted points
(the points which are less attracted by limit values) and \textquotedblleft
penalizes\textquotedblright\ the least interesting ones. When the most
efficient points reach a certain threshold, they are allowed to be multiplied
in their neighborhood. The least efficient points disappear when they reach a
minimal threshold. This system of thresholds which changes the dynamical
structure is our meta-dynamics.

\textbf{Dynamical level (DL):} with the formalism of definition \ref{MAS}, we
have $T_{0}=\mathbb{N}$. The index set $I$ is the set of all finite sets of
points of $\mathbb{C}$ (so $i$ is a set of points of $\mathbb{C}$ and $I$ is
not countable). For all $i\in I$, $X_{i}=\mathbb{N}^{card(i)}$ is the set of
the point weights of $i$. At each given point $z\in\mathbb{C}$, we attach a
selective value
\[%
\begin{array}
[c]{llll}%
\mu_{n}: & \mathbb{C} & \rightarrow & \mathbb{R}^{+}\\
& z & \mapsto & min\left\{  \left\vert P_{c}^{n}(z)\right\vert ,\left\vert
P_{c}^{n}(z)-c\right\vert ,\left\vert P_{c}^{n}(z)-c^{2}-c\right\vert
,\frac{1}{\left\vert P_{c}^{n}(z)\right\vert }\right\}
\end{array}
\]
$\left\{  0,c,c^{2}+c\right\}  $ is the attractive cycle of\ period $3$,
$\frac{1}{P_{c}^{n}(z)}$ is referring to the attraction to the infinity seeing
that $P_{c}^{n}(z)\underset{n\rightarrow+\infty}{\rightarrow}\infty$ is
equivalent to $\frac{1}{P_{c}^{n}(z)}\underset{n\rightarrow+\infty
}{\rightarrow}0$. One increases the weight of the points which are close to
the boundary. In order to do this, we organize a \textquotedblleft
competition\textquotedblright\ by\ comparing their mutual slowness of
convergence. So, the weights of the points $z\in i$ are given by
\begin{align*}
\omega_{z}\left(  t+1\right)   &  =\omega_{z}\left(  t\right)  +\sum_{q\in
i,q\neq z}\delta\left(  z,q\right)  \,\text{with }\delta\left(  z,q\right)
=\left\{
\begin{array}
[c]{ccc}%
1 & \text{if} & \mu_{t}\left(  z\right)  >\mu_{t}\left(  q\right) \\
0 & \text{if} & \mu_{t}\left(  z\right)  =\mu_{t}\left(  q\right) \\
-1 & \text{if} & \mu_{t}\left(  z\right)  <\mu_{t}\left(  q\right)
\end{array}
\right. \\
\omega_{z}\left(  0\right)   &  =0
\end{align*}
where $t\in\mathbb{N}$. Using the notation of section \ref{SecMAS}, we have
the following transition function
\[
\varphi_{i}\left(  t+1,t,\left\{  \omega_{z}\left(  t\right)  \right\}  _{z\in
i}\right)  =\left\{  \omega_{z}\left(  t+1\right)  \right\}  _{z\in i}%
\]
for all $t\in\mathbb{N}$. For each state space there is only one associated
transition function, so the use of the index set $J$ is unnecessary.

\begin{figure}[th]
\centering
\includegraphics[width=7cm]{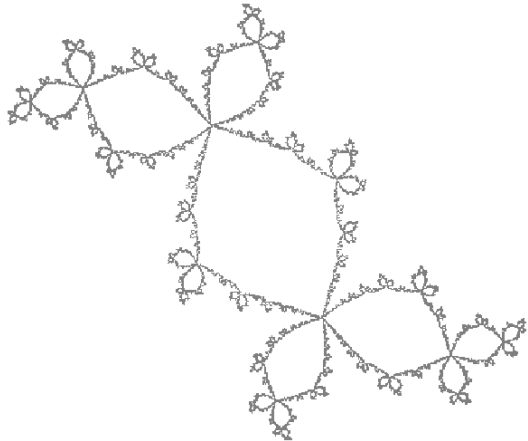}\caption{The Douady's Rabbit fractal}%
\label{Douady}%
\end{figure}

\textbf{Meta-dynamical level (ML):} with the formalism of definition
\ref{MAS}, we have $T_{1}=k\mathbb{N}$. When the weight of a point reach an
upper threshold $M>0$, the point is allowed to give birth to a new point,
randomly in its neighborhood. When the weight of a point reaches a lower
threshold $m<0$, the point is removed. This can be modelled by the
meta-dynamical rule%
\[
\Phi\left(  t,\left\{  \omega_{z}\left(  t\right)  \right\}  _{z\in i\left(
n\right)  },\varphi_{i\left(  n\right)  }\right)  =\left(  \left\{
\omega_{z^{\prime}}\left(  t\right)  \right\}  _{z^{\prime}\in i\left(
n+1\right)  },\varphi_{i\left(  n+1\right)  }\right)
\]
where
\begin{align*}
i\left(  n+1\right)   &  =\left\{  z+\varrho\epsilon:\varrho\text{ is a random
point in }\mathcal{B}\text{, }z\in i\left(  n\right)  \text{ and }\omega
_{z}\left(  t\right)  >M\right\}  \cup\\
&  \left\{  z\in i\left(  n\right)  :\omega_{z}\left(  t\right)  \geq
m\right\}
\end{align*}
such that $\epsilon>0$ given. The weight of the reproduced point is shared
between itself $\omega_{z}\left(  t\right)  $ and the new neighbouring point
$\omega_{z+\varrho\epsilon}\left(  t\right)  $. The other weights are kept
equal. Let us sum up the evolution rule:%
\[
\ldots\rightarrow\varphi_{i\left(  n\right)  }\left(  t\right)  \underset
{\text{DL}}{\rightarrow}\varphi_{i\left(  n\right)  }\left(  t+1\right)
\underset{\text{DL}}{\rightarrow}\ldots\underset{\text{DL}}{\rightarrow
}\varphi_{i\left(  n\right)  }\left(  t+k\right)  \underset{\text{ML}%
}{\rightarrow}\varphi_{i\left(  n+1\right)  }\left(  t+k\right)
\rightarrow\ldots
\]
As the time increases, the selective value becomes more and more accurate,
i.e. $i\left(  n\right)  $ tends to a set of points belonging to the Julia
space $J_{P_{c}}$ or to the empty set when $n\rightarrow+\infty$. This is an
interesting point because the calculations are concentrated on the fractal
boundary. In a lot of classical algorithms, the calculation time is squandered
for points in the interior $\overset{\circ}{K}_{P_{c}}$ (see \cite{Peitgen88}%
). Indeed, this meta-dynamical adaptive system produces a cloud of points
which gather round the Julia space $J_{P_{c}}$ called the Douady's Rabbit
fractal (see Fig. \ref{Douady}).

Though we have developed this algorithm in the setting of Julia sets, the same
framework can be used to explore many complex frontiers, for example other
fractal structures where the boundary properties are largely unknown. Indeed,
this kind of algorithm is really interesting because the calculation is
concentrated on a specific region.

\section{Adaptive differential equations\label{SecADE}}

First, let us recall the classical notion of Hausdorff fractal dimension (see
\cite{Peyriere76}). Let $\mathcal{E}$ be a subspace of a metric space
$\mathcal{M}$ and $\rho$ a positive number, one defines $\mathcal{R}_{\rho}$
as the set of all coverings $\left(  \mathcal{B}_{i},\rho_{i}\right)  _{i}$ of
$\mathcal{E}$ by balls $\mathcal{B}_{i}$ with diameter $0<\rho_{i}<\rho$. For
each positif number $\alpha$, one denotes:
\[
H_{\rho}^{\alpha}\left(  \mathcal{E}\right)  =inf\left\{  \sum\nolimits_{i}%
\rho_{i}^{\alpha}:\left(  \mathcal{B}_{i},\rho_{i}\right)  _{i}\in
\mathcal{R}_{\rho}\right\}  .
\]
$H^{\alpha}\left(  \mathcal{E}\right)  =\lim\limits_{\rho\rightarrow0}H_{\rho
}^{\alpha}\left(  \mathcal{E}\right)  $ is called the $\alpha-$%
\emph{dimensional Hausdorff measure} \emph{of }$\mathcal{E}$ and belongs to
$\left[  0,+\infty\right]  $. Then, let
\[
dim(\mathcal{E})=inf\left\{  \alpha>0:H^{\alpha}\left(  \mathcal{E}\right)
=0\right\}  ,
\]
it is the \emph{Hausdorff dimension} of $\mathcal{E}$. On the one hand, the
Hausdorff dimension is defined for all metric spaces. On the other hand, in
the case of a classical space (non fractal), it is identical to its
topological dimension (for example the Hausdorff dimension of $\mathbb{R}^{n}$
is $n$). In the case of a simple linear fractal, such as fractals with
internal homothetia obtained by an homothetic iteration with constant ratio,
the Hausdorff dimension is equal to the homothetic dimension $dim_{h}\left(
\mathcal{E}\right)  $ given by:
\[
dim_{h}\left(  \mathcal{E}\right)  =\frac{ln(n)}{ln(k)}=log_{k}(n)
\]
where $n$ is the number of subsets obtained during the homothetic process of
reduction with ratio $\frac{1}{k}$ (see \cite{Besicovitch37}). For more
information on dimension theory the reader may refer to \cite{Tricot81} or
\cite{Hurewicz41}.

The variable dimension space is defined as follows:

\begin{definition}
\label{Def EspDimVar}Let $\mathcal{M}$ be a metric space and $\Lambda$ a
parameter space. One defines two maps $d:\Lambda\rightarrow\left[
0,+\infty\right]  ,\lambda\mapsto d(\lambda)$ and $F:\Lambda\rightarrow
2^{\mathcal{M}}$ where $F(\lambda)\ $verifies $dimF(\lambda)=d(\lambda)$ and
$2^{\mathcal{M}}$ is the family of non-empty subsets of $\mathcal{M}$.
$F(\Lambda)$ is a set of variable dimension spaces. If $F(\Lambda)$ is a
totally ordered set for the inclusion, one calls $(F,d)$ a \emph{variable
dimension space}. $d$ is the \emph{dimension function }and $d(F(\lambda))\leq
d(\mathcal{M})$ for all $\lambda\in\Lambda$. If $\Lambda$ is a topological
space then $F$ is a set valued function and we may take account of the
regularity of $d$. The variable dimension space $(F,d)$ is $\emph{continuous}$
if $d$ is continuous. Moreover, if $\Lambda$ is an ordered space then $(F,d)$
is $\emph{increasing}$ (respectively \emph{decreasing}) if $d$ is increasing
(respectively decreasing).
\end{definition}

To illustrate these definitions, we can consider the following example:

\begin{example}
Let $\frac{1}{4}\leq\lambda\leq\frac{1}{2}$, one defines four similarities
from the family $K^{c}$ of the compact subset of the square $c=\left[
0,1\right]  ^{2}$ with value in $K^{c}$ by

\begin{itemize}
\item $s_{1,\lambda}\left(  x\right)  =\lambda x$,
\end{itemize}

\begin{itemize}
\item $s_{2,\lambda}\left(  x\right)  =\lambda\left(
\begin{array}
[c]{ll}%
\frac{1}{2}-\lambda & -\sqrt{\lambda-\frac{1}{4}}\\
\sqrt{\lambda-\frac{1}{4}} & \frac{1}{2}-\lambda
\end{array}
\right)  x+\left(
\begin{array}
[c]{l}%
\lambda\\
0
\end{array}
\right)  $,
\end{itemize}

\begin{itemize}
\item $s_{3,\lambda}\left(  x\right)  =\lambda\left(
\begin{array}
[c]{ll}%
\frac{1}{2}-\lambda & \sqrt{\lambda-\frac{1}{4}}\\
-\sqrt{\lambda-\frac{1}{4}} & \frac{1}{2}-\lambda
\end{array}
\right)  x+\left(
\begin{array}
[c]{c}%
\frac{1}{2}\\
\sqrt{\lambda-\frac{1}{4}}%
\end{array}
\right)  $ and
\end{itemize}

\begin{itemize}
\item $s_{4,\lambda}\left(  x\right)  =\lambda x+\left(
\begin{array}
[c]{c}%
1-\lambda\\
0
\end{array}
\right)  $.
\end{itemize}

Then, one defines the function $\Omega_{\lambda}:K^{c}\rightarrow K^{c}$,
$x\mapsto s_{1,\lambda}\left(  x\right)  \cup s_{2,\lambda}\left(  x\right)
\cup s_{3,\lambda}\left(  x\right)  \cup s_{4,\lambda}\left(  x\right)  $.

\begin{figure}[ptbh]
\centering
\includegraphics[width=12cm]{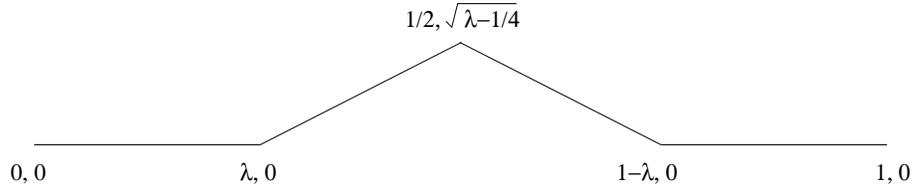}\caption{Van Koch snowflake
generator}%
\label{Koch}%
\end{figure}$\Omega_{\lambda}$ being contracting, one defines $F\left(
\lambda\right)  $ as the fixed point\footnote{To prove this, one uses two
classical results: the family of subsets of a Banach space\ endowed with the
Hausdorff distance is also a Banach space and if the similarities are
contracting in $K^{2}$, then the function $\Omega_{\lambda}$ is also
contracting in $K^{c}$ with the Hausdorff distance \cite{Tricot93}.} of this
function for the Hausdorff distance. In \cite{Mandelbrot80}, one may find the
following result: the dimension of $F\left(  \lambda\right)  $ is $d\left(
\lambda\right)  =\frac{ln4}{ln\left(  \frac{1}{\lambda}\right)  }$ . So,
$\left(  F,d\right)  $ is a continuous increasing variable dimension space on
$\Lambda=\left[  \frac{1}{4},\frac{1}{2}\right]  $. A similar construction
allows us to turn continuously a $n-$dimension space into a $(n+1)-$dimension space.
\end{example}

An interesting application of the above concept arises when the variable
dimension space is the state space of the solutions of a differential equation
over a period of time. The union of the solutions then ``moves'' on a variable
dimension space. We talk about an adaptive differential equation.

\begin{definition}
\label{Adaptive ODE}Let $X$ be a Banach space of finite dimension called the
\emph{possible state space}. Consider

\begin{enumerate}
\item a subdivision $\left\{  t_{i}\right\}  _{i\in\mathbb{N}}$ of
$\mathbb{R}_{+}$,

\item an application $d:\mathbb{R}_{+}\times X\rightarrow\mathbb{N}$, $\left(
t,y\right)  \mapsto d\left(  t,y\right)  $ such that $d\left(  0,y\right)  $
is given,

\item an application $g:\mathbb{R}_{+}\times\mathbb{R}^{dim\left(  X\right)
}\rightarrow\mathbb{R}^{dim\left(  X\right)  }$,
\end{enumerate}

such that one has the system
\begin{equation}
\left\{
\begin{array}
[c]{l}%
\dot{x}_{i}(t)=f_{i}(t,x_{i}(t)),t\in\left[  t_{i},t_{i+1}\right[
,x_{i}(t)\in\mathbb{R}^{d\left(  t_{i},y_{i}\right)  },i\in\mathbb{N}\\
x_{i}\left(  t_{i}\right)  =g\left(  t_{i},y_{i}\right)
\end{array}
\right.  \label{Syst}%
\end{equation}
where $y_{0}=x_{0}\left(  0\right)  $, $y_{i}=\lim\limits_{t\rightarrow
t_{i}^{-}}x_{i-1}\left(  t\right)  $ for $i\in\mathbb{N}^{\ast}$, $\dot{x}%
_{i}\left(  .\right)  $ is the right derivative of $x_{i}\left(  .\right)  $
and
\[
\left\{  f_{i}\right\}  _{i\in\mathbb{N}}:\left[  t_{i},t_{i+1}\right[
\times\mathbb{R}^{d\left(  t_{i},y_{i}\right)  }\rightarrow\mathbb{R}%
^{d\left(  t_{i},y_{i}\right)  }%
\]
is a family of applications. Such a system is called an \emph{adaptive
differential equation}. A \emph{trajectory} of the system (\ref{Syst}) is a
family
\[
x\left(  t\right)  =\left\{  x_{i}\left(  t\right)  :t\in\left[  t_{i}%
,t_{i+1}\right[  \right\}  _{i\in\mathbb{N}}.
\]

\end{definition}

One may notice that $dim(g\left(  t_{i},y_{i}\right)  )=d\left(  t_{i}%
,y_{i}\right)  $. This concept of adaptive differential equation is not a
succession of ordinary differential equations because the initial condition of
each system $i$ depends on the system $i-1$. The ordinary differential
equations represent a case where $d$ is constant.\newline The adaptive
differential equation is a reductive approach of the meta-dynamical adaptive
system to a system described by a differential equation whose meta-dynamics is
discrete and where $\dim X<+\infty$.

\section{Application to a biological model\label{Sec BIO}}

Here, we want to model the influence of the bacterium-phage interaction on the
co-evolution of the populations of bacteria and phages. In the following we
will need the definition of the Hamming distance. Let $n\in\mathbb{N}$, the
\emph{Hamming distance }is the function $d_{H}:\left\{  0,1\right\}
^{n}\rightarrow\mathbb{N}$ defined by
\[
d_{H}\left(  s_{1},s_{2}\right)  =\sum\limits_{k=0}^{n_{c}}\left\vert
s_{i}^{k}-s_{j}^{k}\right\vert
\]
where $s_{1},s_{2}\in\left\{  0,1\right\}  ^{n}$. $d_{H}$ represents the
number of differing bits between the two binary strings $s_{1}=s_{1}^{1}%
s_{1}^{2}\ldots s_{1}^{n}$ and $s_{2}=s_{2}^{1}s_{2}^{2}\ldots s_{2}^{n}$.

The attack of a bacteria population by phages is assumed to be done by the
lysis process: a phage hangs on the surface of a bacterium cell, injects its
DNA in it and then forces the bacterium to yield its own replicas inside the
cell. When the cell is full, it bursts, releasing a huge quantity of copies of
the infecting phage. The efficiency of the attack (i.e. the probability of
success of the infection), depends on the couple bacterium-phage. One of the
other characteristics of bacterial and phages populations are their high
variabilities. They frequently mutate, creating new populations with new
properties. Such a system has two dynamics to be taken into account: the
dynamics of the populations of bacteria and phages and the meta-dynamics of
evolution geared by mutations and extinctions. It is without any doubt a model
which is not in the scope of the classical theory of dynamical systems, but
our formalism applies well to it.

\textbf{Dynamical level:} it is made up of a set of ordinary differential
equations. Let us consider the following system%
\begin{equation}
\{S,\{B_{i}\}_{0\leq i\leq n_{b}},\{P_{j},I_{1,j},I_{2,j},I_{3,j}\}_{0\leq
j\leq n_{p}}\}, \label{Syst-general}%
\end{equation}
with $S$ the concentration of nutrient, $B_{i}$ the different bacteria strains
with $0\leq i\leq n_{b}$ and $P_{j}$ and $I_{k,j}$ the different phages
strains with $0\leq j\leq n_{p}$. To each population $B_{i}$ and $P_{j}$ is
associated a binary string $s_{i}^{b}\in\left\{  0,1\right\}  ^{n}$ and
$s_{i}^{p}\in\left\{  0,1\right\}  ^{n}$. These binary strings code for the
properties of attack (for the phages) or defence (for the bacteria) facing the
infection (see \cite{Baguelin03} for the biological discussion). The
populations are then characterized by the different concentrations and the two
lists of bit strings. The set $L$ of all the pairs of lists is taken as our
second index (the one called $J$ in definition \ref{MAS}).

The model is described by a modified version of Mosekilde equations (see
\cite{Mosekilde96}). This set of ordinary differential equations describes the
interactions of bacterial populations $B_{i}$ and phage populations $P_{j}$ in
a chemostat. $B_{i}$ and $P_{j}$ also symbolize the concentration of these
populations and $S$ the concentration of the nutrient. The process of
infection of bacteria by phages is modelled by three infection stages
$I_{1,j}$, $I_{2,j}$ and $I_{3,j}$. One associates with the ecosystem
(\ref{Syst-general}) the state space $X_{1+n_{b}+4n_{p}}=\mathbb{R}%
_{+}^{1+n_{b}+4n_{p}}$. Over a period of time without appearance or
disappearance of any strain of bacteria or phages, the dynamical evolution of
the system (\ref{Syst-general}) is modelled by the evolution of the
concentrations of $S$, the different bacteria $B_{i}$ and phages $P_{j}$, and
three infection stages $I_{k,j}$. With the formalism of definition \ref{MAS},
that means that $T_{0}=\mathbb{R}$ and for a fixed $l\in L$ the transition
function $\varphi_{1+n_{b}+4n_{p},l}$ is the integration of the following set
of differential equations
\begin{equation}
\left\{
\begin{array}
[c]{l}%
\frac{dB_{i}}{dt}=\frac{\nu SB_{i}}{\kappa+S}-B_{i}\sum\limits_{j=1}^{n_{p}%
}\alpha\omega_{ij}P_{j}-\rho B_{i}\\
\frac{dI_{1,j}}{dt}=P_{j}\sum\limits_{i=1}^{n_{b}}\alpha\omega_{ij}%
B_{i}-3\frac{I_{1,j}}{\tau}-\rho I_{1,j}\\
\vspace{0.2cm}\frac{dI_{2,j}}{dt}=\frac{3}{\tau}(I_{1,j}-I_{2,j})-\rho
I_{2,j}\\
\frac{dI_{3,j}}{dt}=\frac{3}{\tau}(I_{2,j}-I_{3,j})-\rho I_{3,j}\\
\frac{dP_{j}}{dt}=3\frac{\beta I_{3,j}}{\tau}-P_{j}\left(  \sum\limits_{i=1}%
^{n_{b}}\alpha B_{i}+\sum\limits_{j=1}^{n_{p}}\sum\limits_{k=1}^{3}\alpha
I_{k,j}\right)  -\rho P_{j}\\
\frac{dS}{dt}=\rho(\sigma-S)-\sum\limits_{i=1}^{n_{b}}\frac{\nu\gamma SB_{i}%
}{\kappa+S}%
\end{array}
\right.  \label{Syst-bact}%
\end{equation}

with $0\leq i\leq n_{b}$, $0\leq j\leq n_{p}$, $\rho$ the rate of dilution
($\rho=0.0045$ $min^{-1}$), $\kappa$and $\nu$ respectively the saturation term
and the growth from the Monod equation ($\kappa=10\mu g.ml^{-1}$, $\nu=0.024$
$min^{-1}$), $\alpha$ the theoretical adsorption constant depending on phage
and bacterium size ($\alpha=10^{-9}ml.$ $min^{-1}$), $\tau$ a time constant
($\tau=30$ $min$), $\beta$ the number of copies of phage $j$ released during
the burst of the infected bacterial cell ($\beta=100$), $\sigma$ the
continuous supply of substrate ($\sigma=10\mu g.ml^{-1}$), $\gamma$ the amount
of nutrient consumed in each cellular division ($\gamma=0.01ng$) and finally
$\omega_{ij}$ the probability of infection of $B_{i}$ by $P_{j}$ which depends
on the similarity between bit string $s_{i}^{b}$ (attached to bacterium
population $B_{i}$) and $s_{j}^{p}$ (attached to phage population $P_{j}$) as
follows
\[
\omega_{ij}=\left(  1-\frac{d_{H}(s_{i}^{b},s_{j}^{p})}{n_{c}}\right)  ^{2}%
\]
with $n_{c}$ the size of the binary string.\newline Here, we are not
interested in the identification of the biological dynamical level which can
be found in \cite{Baguelin03}. With the given size of a binary string $n_{c}$,
there exists a finite number of possible dynamical systems, here differential
equations. Indeed, the number of possible different populations of bacteria is
equal to the number of parts of the set of binary strings of size $n_{c}$. So,
there are $2^{2^{n_{c}}}$ possible populations of bacteria. For the same
reason, one deduces that the possible number of populations of phages is the
same and thus the total number of possible state spaces is $2^{2^{n_{c}}%
}\times2^{2^{n_{c}}}=2^{2^{n_{c}}+1}$. So, with the notation of definition
\ref{MAS}, it means that $card\left(  J\right)  =2^{2^{n_{c}}+1}$. If we take
for example $n_{c}=10$, one has $2^{1025}\simeq3.6\times10^{308}$ possible
state spaces. Theoretically, one can consider a $3.6\times10^{308}$
dimensional space to embed the system (\ref{Syst-bact}). Nevertheless, when it
comes to numerical simulations, such a big system is impossible to deal with.
One thus understands the need of an adaptive system to describe the system
(\ref{Syst-bact}).

\textbf{Meta-dynamical level (meta-dynamical adaptive system point of view):}
this is the main difference with the model of Mosekilde given in \cite[Chapter
A]{Mosekilde96} which is not evolutive.

\begin{proposition}
\label{Theorem2}Consider a small interval of time $\Delta t$, the adaptive
changes of the system (\ref{Syst-general}) are given by the following
mechanisms
\begin{equation}
p\Psi(t+\Delta t,t,\delta,\varphi_{2+n_{b}+4n_{p},l_{1}})\cdot\left(
\theta_{1},\ldots,\theta_{k},\varphi_{1+k_{\delta}+n_{b}+4n_{p},l_{2}}\right)
=e^{-\lambda\left(  t\right)  }\frac{\lambda\left(  t\right)  ^{k}}{k!}
\label{Birth}%
\end{equation}
and%
\begin{equation}
\Psi(t+\Delta t,t,\delta,\varphi_{2+n_{b}+4n_{p},l_{1}},m_{\delta}%
)=\varphi_{2-\epsilon_{\delta}+n_{b}+4n_{p},l_{2}} \label{Dead}%
\end{equation}
where

\begin{enumerate}
\item formula (\ref{Birth}) is the probability that the species $\delta$ (a
bacterium $B$ or a phage $P$) gives birth to $k\geq1$ mutante\ strains
$\theta_{1},\ldots,\theta_{k}$ on $[t,t+\Delta t[$ with $\lambda\left(
t\right)  =\frac{\delta(t)}{\delta_{e}}p_{e}$, $p_{e}$ the probability that a
small group of mutant\ species of size $\delta_{e}$ gradually replacing the
species of the parent population $\delta$.$\ \varphi_{i,j}$ is the function
defined by equations (\ref{Syst-bact}) for $j=l_{1},l_{2}$ and $k_{\delta}=k$
if $\delta$ is a bacterium and $k_{\delta}=4k$ if $\delta$ is a phage,

\item formula (\ref{Dead}) gives the determinist rule of the extinction of the
species $\delta$ which depends on a given threshold $m_{\delta}$,
$\epsilon_{\delta}=1$ if $\delta$ is a bacterium and $\epsilon_{\delta}=4$ if
$\delta$ is a phage.
\end{enumerate}
\end{proposition}

$p\Psi$ is the stochastic transition function of definition \ref{SMAS} which
governs the meta-dynamical rule of the appearance. $\Psi$ is the transition
function of definition \ref{MAS} which governs the meta-dynamical rule of the extinction.

\begin{proof}
Consider the probability $p_{B_{i}}(t,k)$ that the population $B_{i}$ gives
birth to $k$ mutante\ strain on $[t,t+\Delta t[$ and the probability $p_{e}$
that a small group of mutante\ bacteria of size $B_{e}$ is gradually replacing
the bacteria of the parent population $B_{i}$. Such a reasoning gives a
binomial probability for
\[
p_{B_{i}}(t,k)=C_{n\left(  t\right)  }^{k}p_{e}^{k}(1-p_{e})^{n\left(
t\right)  -k}%
\]
with $n\left(  t\right)  =\frac{B_{i}(t)}{B_{e}}$. When $n\left(  t\right)  $
is large, one may approximate the binomial probability by the Poisson
probability
\begin{equation}
p_{B_{i}}(t,k)\simeq e^{-\lambda\left(  t\right)  }\frac{\lambda\left(
t\right)  ^{k}}{k!} \label{eq_proba}%
\end{equation}
with $\lambda\left(  t\right)  =\frac{B_{i}(t)}{B_{e}}p_{e}$. Suppose that the
birth of all the populations to a Hamming distance of one (only one bit is
different) is equiprobable. The birth of a mutant strain results in the change
of a group of bacteria\ (which is a part of a parent population) of size
$B_{e}$, in a population group with new characteristics. If this population
group already exists, the mutant population is added to it. The same mechanism
governs the modelling of the phage mutation. For the extension, there exists a
threshold $m_{\delta}$ below which the extinction of the population is
certain. Every population under a given threshold (different for bacteria
$m_{B}$ and phages $m_{P}$) is removed from the system. Thus,the extinction is determinist.
\end{proof}

We have defined the macroscopic birth of a mutant population as an event
occurring on an interval $[t,t+\Delta t[$. There exists a time set $T_{1}$
where the system (\ref{Syst-bact}) commutes. This commutation depends on the
state of the system (\ref{Syst-bact}) and on the coefficient $p_{B}$ et
$p_{P}$ respectively coefficient of the Poisson law of the bacterium and the
phage. In order to simplify the model and to make it computable, one may
suppose that
\[
T_{1}=t_{0}+i\Delta t
\]
even if in practice, the mutations have no reasons to be periodically defined.
This allows to define, with the previous notations, the meta-dynamical rule%
\begin{align*}
p\Phi(t_{0}+i\Delta t,\delta,\varphi_{2+n_{b}+4n_{p},l_{1}})\cdot\left(
\theta_{1},\ldots,\theta_{k},\varphi_{1+k_{\delta}+n_{b}+4n_{p},l_{2}}\right)
&  =e^{-\lambda\left(  t\right)  }\frac{\lambda\left(  t\right)  ^{k}}{k!}\\
\Phi(t_{0}+i\Delta t,\delta,\varphi_{2+n_{b}+4n_{p},1},m_{\delta})  &
=\varphi_{2-\epsilon_{\delta}+n_{b}+4n_{p},2}.
\end{align*}
Then, at each step $\Delta t$ there are four possible commutations:

\begin{itemize}
\item birth of a new bacterial strain: a variable $B_{n_{b}+1}$ is added, the
dimension of the system (\ref{Syst-bact}) increases by one,

\item birth of a new phagical strain: four variables are added: $P_{n_{p}+1}
$, $I_{1,n_{p}+1}$, $I_{2,n_{p}+1}$ and $I_{3,n_{p}+1}$, the dimension of the
system (\ref{Syst-bact}) increases by four,

\item extinction of a bacterial strain: the concerned variable is removed, the
dimension of the system (\ref{Syst-bact}) decreases by one,

\item extinction of phagical strain: variables of the concerned phage are
removed, the dimension of the system (\ref{Syst-bact}) decreases by four.
\end{itemize}

\begin{figure}[ptbh]
\centering
\includegraphics[width=10cm]{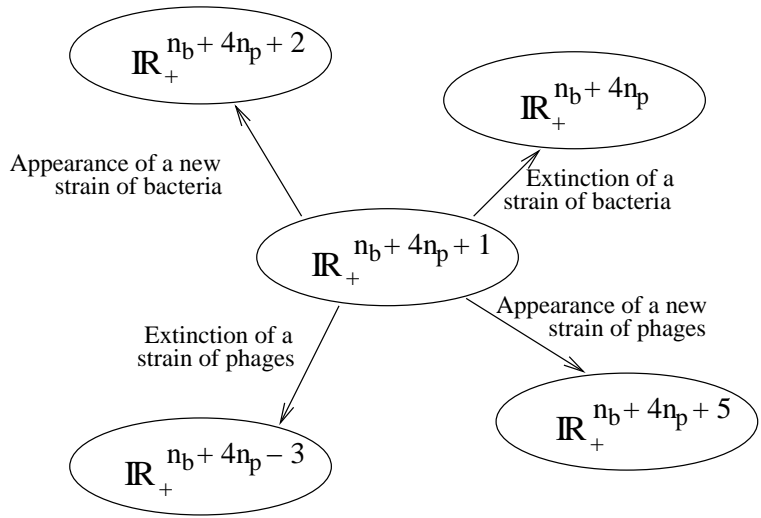}\caption{Evolution of the state
space with $n_{b}$ and $n_{p}$ respectively the number of bacterial and phage
strains before the transition.}%
\label{fig:comutmut}%
\end{figure}

The different possible state spaces resulting from a commutation are given on
figure \ref{fig:comutmut}.

If more than one event occurs at each moment $t\in T_{1}$, one
composes the possible change of state spaces (for example, the extinction of a
phage combined with the birth of a bacterial strain decreases the dimension of
the system (\ref{Syst-bact}) by $4-1=3$).

\textbf{Meta-dynamical level (adaptive differential equation point of view):
}we have described the biological model as a stochastic meta-dynamical
adaptive system by using definition \ref{SMAS}. This modelling corresponds to
a stochastic view of the system (\ref{Syst-general}) where the space of all
stochastic realizations is infinite. There exists a family of points $\left\{
t_{i}\right\}  _{i\in\mathbb{N}}$ where the system (\ref{Syst-bact}) commutes.
We have supposed that this commutation is given by $t_{i}=t_{0}+i\Delta t$.
Then, at each step $\Delta t$ the system may commute. To see a possible
evolution of the system (\ref{Syst-general}), we choose a stochastic
realization at each point $t_{i}$ (only for the appearance of a strain because
the extinction of a strain is determinist) $g(t_{i},y_{i})$ where $y_{i}%
=\lim\limits_{t\rightarrow t_{i}^{-}}x_{i-1}\left(  t\right)  $ with
$x_{i-1}(t)$ the solution of the system (\ref{Syst-bact}) on $\left[
t_{i-1},t_{i}\right[  $. It describes the determinist evolution of the system
(\ref{Syst-general}). By extension, for us a \emph{stochastic realization} is
a function
\[
g:\mathbb{R}_{+}\times\mathbb{R}^{dim\left(  X\right)  }\rightarrow
\mathbb{R}^{dim\left(  X\right)  }%
\]
defined at least on $\left\{  (t_{i},y_{i})\right\}  _{i\in\mathbb{N}}$. For a
given realization $g$, the system (\ref{Syst-general}) may be modelled by an
adaptive differential equation whose equations are given by the system
(\ref{Syst-bact}). This modelling belongs to the variable dimensional space
$\mathbb{R}_{+}^{n_{b}+4n_{p}+1}$ where $n_{b}+4n_{p}+1$ depends on $g$ and
follows the rule given on figure \ref{fig:comutmut}. A detailed study of this
biological model with implementation can be found in \cite{Baguelin02}.

In this example, one sees that the framework of the adaptive differential
equations corresponds to the case of a transition function defined by a set of
differential equations and a meta-dynamical time reduced to a discrete subset
of $\mathbb{R}$.

\section{Conclusion}

We started our research with a biological system whose dynamics changes in
different dimensions. Seeing that there was no mathematical framework to
describe such a system, we have developed a mathematical tool following
Kalman's dynamical system called meta-dynamical adaptive system which was
appropriate to give a constructive algorithm for some fractals. It was also
adapted to describe and analyze our biological system. However, we have found
our tool too general and we decided to develop a special tool for differential
equations. The new system called \textquotedblleft adaptive differential
equation\textquotedblright\ is not a succession of differential equations
because the initial condition of each system depends on the previous system
and gives the new dimension of the following system. This model allows to
describe a system of changing differential equations, in particular the
stochastic realization of a stochastic meta-dynamical adaptive system. The
last tool we use is the \textquotedblleft variable dimension
space\textquotedblright. This new kind of space links the notion of space and
dimension in a changing dynamics. We think that our work will contribute to
understand the huge number of complex systems where the espace of exploration
is too big to be investigate with classical means.

\section*{Acknowledgement}

The authors would like to thank Jacques LeF\`{e}vre to have given a start to
our study of evolutionary ecosystems and Jean-Pierre Richard at the head of
the SyNeR team for his help in producing this article.

\bibliographystyle{plain}
\bibliography{emmanuel}

\end{document}